\documentclass[twocolumn,showpacs,pra,aps,floatfix]{revtex4}

\usepackage{graphicx}
\usepackage{bm}
\usepackage{psfrag}
\usepackage{amsmath}
\usepackage{amssymb}
\usepackage{latexsym}
\usepackage{exscale}

\newcommand{\ten}[1]{\mbox{\textbf{\textsf{#1}}}}

\begin{document}

\title{Local-field corrected van der Waals potentials
in magnetodielectric multilayer systems
}

\author{Agnes Sambale}
\author{Dirk-Gunnar Welsch}
\affiliation{Theoretisch-Physikalisches Institut,
Friedrich-Schiller-Universit\"at Jena,
Max-Wien-Platz 1, D-07743 Jena, Germany}

\author{Ho Trung Dung}
\affiliation{Institute of Physics, Academy of
Sciences and Technology, 1 Mac Dinh Chi Street,
District 1, Ho Chi Minh city, Vietnam}

\author{Stefan Yoshi Buhmann}
\affiliation{Quantum Optics and Laser Science, Blackett Laboratory,
Imperial College London, Prince Consort Road,
London SW7 2AZ, United Kingdom}

\date{\today}

\begin{abstract}
Within the framework of macroscopic quantum electrodynamics in linear, 
causal media, we study the van der Waals potentials of ground-state
atoms in planar magnetodielectric host media. Our investigation extends
earlier ones in two aspects: It allows for the atom to be embedded in
a medium, thus covers many more realistic systems; and it takes
account of the local-field correction. Two- and three-layer
configurations are treated in detail both analytically and
numerically. It is shown that an interplay of electric and magnetic
properties in neighbouring media may give rise to potential wells or
walls. Local-field corrections as high as 80\% are found. By
calculating the full potential including the translationally invariant
and variant parts, we propose a way to estimate the (finite) value of
the dispersion potential at the surface between two media. Connection
with earlier work intended for biological applications is established.
\end{abstract}

\maketitle

%%%%%%%%%%%%%%%%%%%%%%%%%%%%%%%%%%%%%%%%%%%%%%%%%%%%%%%%%%%%%%%%%%%%%%

\section{Introduction}
\label{sec_intro}

The dispersive interaction between small neutral, unpolarised
particles (atoms, molecules etc.) and an uncharged macroscopic 
object is a manifestation of the quantum nature of the electromagnetic
field and related to the zero-point energy \cite{Review2006}. In the
first quantum electrodynamical treatment of this van der Waals (vdW)
type force \cite{Casimir}, and in many other related works, the atom
was assumed to be located in free space. In reality, the atom may be
submerged in or be a part of a gas, liquid, or even solid --- a
situation typical in diverse fields such as colloid science
\cite{Tadros1993,Gregory1993,Thomas1999}, surface engineering
\cite{Lubarsky2006}, and biology \cite{Nir1976}. As a particular
example from biology, one can cite the transfer of a small particle
diluted in cell plasma through a cell membrane caused by vdW forces
\cite{Israelachvili1974}. A question that may arise when the atom is
embedded in a medium is how the vdW interaction is to be corrected due
to the difference between the local fields acting on the atom and the
macroscopic fields averaged over a region which contains a great
number of the medium constituents, thus ignoring the gaps between
them. In Ref.~\cite{Tomas2007}, the  vdW interaction between two
ground-state atoms embedded in adjacent semi-infinite
magnetodielectric media has been considered. Comparison of this result
with that deduced from the Casimir force on a thin composite slab in
front of a composite semi-infinite medium, both obeying the
Clausius-Mossotti relation, suggests a hint on how to account for the
local-field correction. This confirmed shortly after in
Ref.~\cite{Sambale2007a} on the basis on a macroscopic quantum
electrodynamics theory and the real-cavity model~\cite{Onsager1936}.
A general formula for the vdW potential in an arbitrary geometry has
been derived in the form of a sum of a translationally invariant term
and and a term containing the uncorrected scattering Green tensor modified
by a local-field correction factor \cite{Tomas2007,Sambale2007a}. This
general formula is applicable for meta-materials \cite{Lezec2007} and
will serve as a starting point for our treatment of the local-field
corrected vdW interaction in a stratified magnetodielectric.
Note that a generalisation of the formalism has recently been used to study 
local-field corrected interactions of an excited atom with a ground-state one 
across an interface between two media \cite{Tomas07}.

The interaction between a neutral atom and a material surface is
customarily split into two parts: a short-range repulsive part,
significant when the atomic valence electrons overlap with the
surface, and a longer-range dispersive vdW part \cite{Kleiman1973,
Zaremba1977, Nordlander1984, Holmberg1984,Das1986, Holmberg1986,
Mukhopadhyay1987}. Theories that focus on the vdW interaction commonly
neglect the first, thus are incapable of correctly predicting the
behaviour of the interaction potential at very short distances. They
typically yield a power law of $z_A^{-3}$ ($z_A$, atom--surface
distance) for materials with dominant electric properties, and a power
law of $z_A^{-1}$ for purely magnetic materials, which lead to
divergent values for the vdW potential right on the surface
\cite{Casimir, Lifshitz1955, Kampf2005}. Due to the importance of
phenomena such as physisorption and transport of particles through
interfaces and membranes, much effort has been devoted to a better
treatment of the atom--surface interaction when the two are at very
short distances. With respect to physisorption, this has been done via
introducing a reference plane \cite{Zaremba1976}; via characterising
the material surface by a more realistic response function which
includes spatial dispersion \cite{Mukhopadhyay1975}, smooth variation
of the dielectric properties at the interface, and the contribution
from d electrons to the screening \cite{Holmberg1984a}; and via using
an atomic polarisability going beyond the dipole approximation
\cite{Holmberg1984a, Holmberg1984}. These studies typically produce
finite values for the interaction potential at the interface
\cite{Zaremba1976}.

In the present paper, the local-field corrected vdW interaction of a 
ground-state atom embedded in a planar, dispersing and absorbing
magnetoelectric host medium is studied within the framework of
macroscopic quantum electrodynamics and the real-cavity model. We
propose a procedure that would allow for obtaining an estimate of
the finite vdW energy of an atom right on the interface even within
our predominantly macroscopic framework treating the atom within
electric dipole approximation. The paper is organised as follows. In
Sec.~\ref{sec_real} the main results concerning the local-field
corrected vdW interaction of a ground-state atom and an arbitrary
absorbing and dispersing macroscopic body are reviewed, and then
applied to planar multilayer systems. The two-layer case is considered
and a procedure for an estimation of the vdW potential at the
interface is given in Sec. \ref{sec_twolayer} while the three-layer
case is discussed in Sec. \ref{subsec_three}. Connection with earlier
work which focus on biological systems and concluding remarks are
given in Sec.~\ref{sum}.

%%%%%%%%%%%%%%%%%%%%%%%%%%%%%%%%%%%%%%%%%%%%%%%%%%%%%%%%%%%%%%%%%%%%%%

\section{The Model}
\label{sec_real}

We set the stage by briefly recalling the results for the local-field
corrected vdW potential of a ground-state atom within an arbitrary
geometry and applying them to a planar multilayer system.

\subsection{Local-field corrected vdW potentials in arbitrary
geometries}

We consider a ground-state guest atom $A$ located at $\mathbf{r}_A$ in
an absorbing and dispersing host medium of arbitrary size and shape
characterised by macroscopic $\varepsilon(\mathbf{r},\omega)$ and
$\mu(\mathbf{r},\omega)$. To account for the local-field correction we
employ the real-cavity model by assuming the atom to be surrounded by
a small spherical free-space cavity of radius $R_\mathrm{c}$. The
radius of the cavity is a measure of the distance between the guest
atom and the surrounding host atoms \cite{Sambale2007a}. By
construction, the macroscopic quantities
$\varepsilon(\mathbf{r},\omega)$ and $\mu(\mathbf{r},\omega)$ do not
vary appreciably on the microscopic length scale $R_\mathrm{c}$. To
apply the real cavity model one should keep in mind that
$\sqrt{\varepsilon(0)\mu(0)}R_\mathrm{c}$ should be small compared to
the maximum of all characteristic atomic and medium wavelengths as
well as to the separation from any surface of the host medium. In
particular, the application of model to metals is very problematic.
Using second-order perturbation theory, the vdW potential for such an
atom can be written in the form \cite{Sambale2007a}
\begin{equation}
U(\mathbf{r}_A)=U_1(\mathbf{r}_A)+U_2(\mathbf{r}_A),
\end{equation}
where $U_1(\mathbf{r}_A)$ is constant throughout any homogenous
region and accounts for all scattering processes within the cavity, 
\begin{equation}
\label{U1}
U_1(\mathbf{r}_A)=-\frac{\hbar\mu_0}{4\pi^2 c}
\int _0^{\infty}\mathrm{d} \xi\, \xi^3 \alpha(i\xi) C_A(i\xi),
\end{equation}
where
\begin{multline}
\label{C}
C_A(\omega)=\\
\frac{h_1^{(1)}(z_0)\left[zh_1^{(1)}(z)\right]'
-\varepsilon_A(\omega)
h_1^{(1)}(z)\left[z_0 h_1^{(1)}(z_0)\right]'}
{\varepsilon_A(\omega)
h_1^{(1)}(z)\left[z_0 j_1(z_0)\right]'-j_1(z_0)
\left[zh_1^{(1)}(z)\right]'}
\end{multline}
[$z_0$ $\!=$ $\!\omega R_\mathrm{c}/c$, $z$ $\!=$ $\!n(\omega) z_0$,
$n(\omega)$ $\!=$ $\!\sqrt{\varepsilon_A(\omega) \mu_A
(\omega)}$, $\varepsilon_A(\omega)=\varepsilon(\mathbf{r}_A,\omega)$, 
$\mu_A(\omega)=\mu(\mathbf{r}_A, \omega)$], with $j_1(x)$ and
$h_1^{(1)}(x)$, respectively, being the first spherical Bessel
function and the first spherical Hankel function of the first kind,
\begin{align}
&j_1(x)=\frac{\sin(x)}{x^2}-\frac{\cos(x)}{x},\\
&h_1^{(1)}(x)=-\left(\frac{1}{x}+\frac{i}{x^2}\right)e^{ix},
\end{align}
and $\alpha(\omega)$ denoting the isotropic polarisability of the
guest atom in lowest non-vanishing order of perturbation theory
\cite{fain},
\begin{equation}
\label{alpha}
\alpha(\omega)=\lim _{\eta\rightarrow 0+}\frac{2}{3\hbar} \sum _k
\frac{\omega_{k0}\left|\mathbf{d}^{0k}\right|^2}
{\omega_{k0}^2-\omega^2-i\eta\omega}
\end{equation}
with $\omega_{k0}=(E_k-E_0)/\hbar$ being the (unperturbed) atomic
transition frequencies and $\mathbf{d}^{lk}\equiv\langle
k|\hat{\mathbf{d}}|l\rangle$ being the atomic electric-dipole
transition matrix elements. 

The term $U_2(\mathbf{r}_A)$ involves all interactions associated with
the particular shape and size of the magnetodielectric host medium,  
\begin{equation}
\label{U2}
U_2(\mathbf{r}_A)\!=\!\frac{\hbar\mu_0}{2\pi}
\int _0^{\infty}\mathrm{d} \xi\, \xi^2
D_A^2(i\xi)\alpha(i\xi)
\mathrm{Tr}\,
\ten{G}^{(1)}(\mathbf{r}_A,\mathbf{r}_A,i\xi)
\end{equation}
where
\begin{multline}
D_A(\omega)=\\
\frac{j_1(z_0)\left[z_0h_1^{(1)}(z_0)\right]'
-\left[z_0j_1(z_0)\right]'h_1^{(1)}(z_0)}
{\mu_A(\omega)\Bigl
[j_1(z_0)\left[z h_1^{(1)}(z)\right]'
-\varepsilon_A(\omega)
\left[z_0j_1(z_0)\right]'h_1^{(1)}(z)
\Bigr
]}.
\end{multline}
The Green tensor $\ten{G}^{(1)}(\mathbf{r},\mathbf{r'}, \omega)$ of
the electromagnetic field accounts for scattering at the
inhomogeneities of the (unperturbed) magnetoelectric host medium while
the factor $D_A(\omega)$ comes from the local-field correction. 

Within the real-cavity model, the potentials $U_1(\mathbf{r}_A)$ and
$U_2(\mathbf{r}_A)$ are well approximated by their asymptotic limit
of small cavity radii, i.e., we keep only the leading non-vanishing
order in $\sqrt{|\varepsilon(0)\mu(0)|}\omega_\mathrm{max}
R_\mathrm{c}/c$ where $\omega_{max}$ represents the maximum of
characteristic atomic and medium frequencies (for details, see
Ref.~\cite{Sambale2007a}), 
\begin{multline}
\label{U1approx}
U_1(\mathbf{r}_A)=-\frac{\hbar}{4\pi^2\varepsilon_0}\int _0^{\infty}
\mathrm{d} \xi\,
\alpha\left[3\,\frac{\varepsilon_A-1}{2\varepsilon_A+1}\frac{1}{
R_\mathrm{c}^3 }\right.\\
\left. +\frac{9\xi^2}{c^2}\,
\frac{\varepsilon_A^2\left[1-5\mu_A\right]+3\varepsilon_A+1}
{5\left[2\varepsilon_A+1\right]^2}\frac{1}{R_\mathrm{c}}\right]
\end{multline}
and
\begin{multline}
\label{U2approx}
U_2(\mathbf{r}_A) = \frac{\hbar \mu_0}{2\pi} \int _0 ^{\infty}
\mathrm{d}\xi\,
\xi^2 \alpha\left(\frac{3\varepsilon_A}{2\varepsilon_A+1}\right)^2 \\
\times\mathrm{Tr}\,\ten{G}^{(1)}(\mathbf{r_A},\mathbf{r_A},i\xi).
\end{multline}
Here and in the following the dependence of $\varepsilon$, $\mu$ and
$\alpha$ on the $i\xi$ is suppressed for brevity. Note that in leading
order the local-field factor  $[3\varepsilon_A/(2\varepsilon_A+1)]^2$
depends on dielectric properties only. The associated (conservative)
vdW force is given by 
\begin{multline}
\label{vdwforce2}
\mathbf{F}(\mathbf{r}_A)=-\bm{\nabla} U_2(\mathbf{r}_A)= -\frac{\hbar
\mu_0}{2\pi} \int _0^{\infty} \mathrm{d}\xi\, \xi^2 \alpha\\
\times \left(\frac{3\varepsilon_A}{2\varepsilon_A+1}\right)^2
\bm{\nabla}
\mathrm{Tr}\,\ten{G}^{(1)}(\mathbf{r}_A,\mathbf{r}_A,i\xi).
\end{multline}
The cavity-induced part of the potential, $U_1$ according to 
Eq.~(\ref{U1}) does not lead to a force but to an energy shift whose
influence on the overall potential will be studied in
Secs.~\ref{sec_twolayer} and \ref{subsec_three}.

%%%%%%%%%%%%%%%%%%%%%%%%%%%%%%%%%%%%%%%%%%%%%%%%%%%%%%%%%%%%%%%%%%%%%%

\subsection{Atom in a magnetoelectric multi-layer system}
\label{sec_multi}

So far we have not been specific about the geometry of the macroscopic
body. Consider a stack of $n$ layers labeled by $l$ ($l=1,\dots, n$)
of thicknesses $d_l$ with planar parallel boundary surfaces, where
$\varepsilon(\mathbf{r},\omega)=\varepsilon_l(\omega)$ and
$\mu(\mathbf{r},\omega)=\mu_l(\omega)$ in layer $l$. The coordinate
system is chosen such that the layers are perpendicular to the $z$
axis and extend from $z=0$ to $z=d_l$ for $l\neq 1,n$ and from $z=0$
to $z=-\infty$ ($\infty$) for $l=1$ (n). The scattering part of the
Green tensor at imaginary frequencies for $\mathbf{r}$ and
$\mathbf{r'}$ in layer $j$ is given by (see, e.g.,
Ref.~\cite{Kampf2005})
\begin{equation}
\label{green_int}
\ten{G}^{(1)}(\mathbf{r},\mathbf{r'},i\xi)=\int \mathrm{d}^2 q\, 
e^{i\mathbf{q}\cdot(\mathbf{r}-\mathbf{r'})}
\ten{G}^{(1)}(\mathbf{q},z,z',i\xi)
\end{equation}
[$\mathbf{q}\bot \mathbf{e}_z$] where
\begin{multline}
\label{greenplanar}
\ten{G}^{(1)}(\mathbf{q},z,z',i\xi)\!=\!\frac{\mu_j}{8\pi^2\beta_j}
\!\sum _{\sigma=s,p}\left\{\frac{1}{D_j^\sigma}\right.\\
\times\left[\mathbf{e}_\sigma^+\mathbf{e}_\sigma^-r_{j-}^\sigma 
e^{-\beta_j(z+z')}
+\mathbf{e}_\sigma^-\mathbf{e}_\sigma^+r_{j+}^\sigma 
e^{- 2\beta_j d_j}e^{\beta_j(z+z')}\right]\\
\left.+\frac{r_{j-}^\sigma r_{j+}^\sigma e^{-2\beta_j
d_j}}{D_j^\sigma}\left[\mathbf{e}_\sigma^+ \mathbf{e}_\sigma^+
e^{-\beta_j(z-z')}+\mathbf{e}_\sigma^-\mathbf{e}_\sigma^-
e^{\beta_j(z-z')}\!\right]\!\right\}
\end{multline}
[$j>0$, for $j=0$ set $d_0=0$] with the abbreviation
\begin{equation}
\label{equ7}
D_j^\sigma=1-r_{j-}^\sigma r_{j+}^\sigma e^{-2\beta_jd_j}.
\end{equation}
In Eq.~(\ref{greenplanar}), $p$($s$) denotes $p$($s$) polarisations.
The reflection coefficients obey the recursion relations
\begin{align}
\label{equ8}
r_{l\pm}^\sigma=
\frac{r^\sigma_{ll\pm1}+e^{-2\beta_{l\pm1}d_{l\pm 1}}
r_{l\pm1\pm}^\sigma}{1+r^\sigma_{ll\pm1}r_{l\pm1\pm}^\sigma 
e^{-2\beta_{l\pm1}d_{l\pm 1}}},
\end{align}
\begin{equation}
\label{equ9b}
r^p_{ll+1}=\frac{\varepsilon_{l+1}\beta_l
-\varepsilon_l\beta_{l+1}}{\mu_{l+1}\beta_l+\varepsilon_l\beta_{l+1}},
\; 
r^s_{ll+1}=\frac{\mu_{l+1}\beta_l-\mu_l\beta_{l+1}}{\mu_{l+1}\beta_l
+\mu_l\beta_{l+1}}
\end{equation}
[$r_{1-}^\sigma\!=\!r_{n+}^\sigma\!=\!0$], where the modulus of wave
vector in the $z$ direction is given by
\begin{equation}
\label{equ6b}
\beta_l= \sqrt{k_l^2+q^2}
\end{equation}
with $k_l$, which always appears in the form of $k_l^2$ in the Green
tensor, Eq.~(\ref{greenplanar}), being
\begin{equation}
\label{equ6}
k_l^2=\frac{\xi^2}{c^2}\,\varepsilon_l\mu_l.
\end{equation}
The $s$- and $p$-polarisation unit vectors are defined as 
\begin{equation}
\label{equ5}
\mathbf{e}_s^\pm=\mathbf{e}_q\times\mathbf{e}_z,\quad 
\mathbf{e}_p^\pm=
\frac{1}{i k_j}\,(q\mathbf{e}_z\mp i\beta_j\mathbf{e}_q)
\end{equation}
[$\mathbf{e}_q=\mathbf{q}/q,q=|\mathbf{q}|$]. 
Substitution of Eqs.~(\ref{green_int}) and (\ref{greenplanar}) 
into Eq.~(\ref{U2approx}) leads to
\begin{multline}
\label{Umiddle}
U_2(z_A)=\frac{\hbar\mu_0}{8\pi^2} \int _0^\infty \mathrm{d}\xi\, 
\xi^2\alpha\mu_j
\left[\frac{3\varepsilon_j}{2\varepsilon_j+1}\right]^2
\\
\times\int _0^\infty \!\mathrm{d}q\frac{q}{\beta_j}\! 
\left\{e^{-2\beta_j
z_A}\left[\frac{r_{j-}^s}{D_j^s}\!-\!\left(\!1+2\frac{q^2c^2}{
\xi^2\varepsilon_j\mu_j}\right)\frac{r_{j-}^p}{D_j^p}\right]\right.
\\
\left.+e^{-2\beta_j(d_j-z_A)}\left[\frac{r_{j+}^s}{D_j^s}-
\left(1+2\frac{q^2c^2}{\xi^2\varepsilon_j\mu_j}\right)\frac{r_{j+}^p}{
D_j^p}\right]\right\}\\
+\frac{\hbar\mu_0}{4\pi^2}\int _0^\infty\mathrm{d}\xi \xi^2\alpha 
\mu_j\left[\frac{3\varepsilon_j}{2\varepsilon_j+1}\right]^2
\\
\times \int _0 ^\infty\mathrm{d}q\frac{q}{\beta_j}
\sum _{\sigma=s,p}\frac{r_{j-}^\sigma r_{j+}^\sigma 
e^{-2\beta_j d_j}}{D_j^\sigma},
\end{multline}
where we have used the relations
\begin{align}
\mathbf{e}_s^{\pm}\cdot \mathbf{e}_s^{\pm} =\;& \mathbf{e}_s^{\pm}
\cdot 
\mathbf{e}_s^{\mp}=1,\\
\mathbf{e}_p^{\pm}\cdot \mathbf{e}_p^{\pm} =\;& 1,\quad
\mathbf{e}_p^{\pm}\cdot 
\mathbf{e}_p^{\mp}=-1-2\frac{q^2c^2}{\xi^2 \varepsilon_j\mu_j}
\end{align}
to calculate the trace. It is worth noting that the term in curly
brackets in Eq.~(\ref{Umiddle}) describes processes that involve an
odd number of reflections at the interfaces while the second term
accounts for an even number of reflections, as can be seen from 
\begin{multline}
\label{odd1}
\frac{r_{j-}^\sigma e^{-2\beta_j z_A}}{1-r_{j-}^\sigma 
r_{j+}^\sigma e^{-2\beta_j d_j}}=e^{-\beta_j z_A} 
r_{j-}^\sigma e^{-\beta_j z_A}\\
+e^{-\beta_j z_A} r_{j-}^\sigma e^{-\beta_j d_j} 
r_{j+}^\sigma e^{-\beta_j d_j}r_{j-}^\sigma e^{-\beta_j z_A}+\dots,
\end{multline}
\begin{multline}
\label{odd2}
\frac{r_{j+}^\sigma e^{-2\beta_j (d_j-z_A)}}{1-r_{j-}^\sigma 
r_{j+}^\sigma e^{-2\beta_j d_j}}=e^{-\beta_j (d_j- z_A)} 
r_{j+}^\sigma e^{-\beta_j (d_j-z_A)}\\
+e^{-\beta_j (d_j-z_A)} r_{j+}^\sigma e^{-\beta_j d_j} 
r_{j-}^\sigma e^{-\beta_j (d_j-z_A)}+ \dots
\end{multline}
and
\begin{multline}
\label{even}
\frac{r_{j-}^\sigma r_{j+}^\sigma e^{-2\beta_j d_j}}{1-r_{j-}^\sigma 
r_{j+}^\sigma 
e^{-2 \beta_j d_j}} 
=e^{-\beta_j z_A}r_{j-}^\sigma e^{-\beta_j d_j}r_{j+}^\sigma \\
\times e^{-\beta_j (d_j-z_A)}
+e^{-\beta_j z_A}r_{j-}^\sigma e^{-\beta_j d_j}
r_{j+}^\sigma e^{-\beta_j d_j}\\ 
\times r_{j-}^\sigma e^{-\beta_j d_j}r_{j+}^\sigma 
e^{-\beta_j(d_j-z_A)}+\dots .
\end{multline}
Obviously, the expression presented in Eq.~(\ref{even}), which
corresponds to even numbers of reflections, is independent of the
position of the atom and rapidly decreases with increasing distance
between two plates. Hence, it cannot lead to a force but only to a
layer-dependent energy shift which is small compared to the other
terms. It is completely absent in the two-layer case since only single
reflections at the interface occur. It also vanishes for sufficiently 
dilute media, an expansion for small $\chi_{j}=\varepsilon_j-1$ and
$\zeta_{j}=\mu_j-1$ showing no contribution to linear order, which
reflects the fact that at least three medium-assisted reflection
processes are involved. On the contrary, Eqs.~(\ref{odd1}) and
(\ref{odd2}) depend on the position of the atom, where waves
propagating to the left and to the right of the atom have to be
distinguished. 

It is worth noting that all formulas in this section are valid for
arbitrary (passive) magnetoelectric media, including metamaterials and
in particular, lefthanded materials with simultaneously negative real
parts of $\varepsilon$ and $\mu$, due to the fact that
$\sqrt{\varepsilon_j(\omega) \mu_j(\omega)}$ has no branching points
in the upper half of the complex frequency plane. However, left-handed
material properties being only realized in finite frequency windows,
they can not be expected to have a strong influence on ground-state
dispersion potentials which depend on the medium response at all
frequencies; alternatively, this can be seen from the fact that the
potential is expressible in terms of permittivities and permeabilities
taken at imaginary frequencies which are always positive
\cite{Kampf2005, Spagnolo2007}. The situation can change when the atom
is in an excited state and the atom-field interaction resonantly
depends on the medium response at the atomic transition frequencies
\cite{Sambale2008a, Sambale2008b}.

%%%%%%%%%%%%%%%%%%%%%%%%%%%%%%%%%%%%%%%%%%%%%%%%%%%%%%%%%%%%%%%%%%%%%%
%%%%%%%%%%%%%%%%%%%%%%%%%%%%%%%%%%%%%%%%%%%%%%%%%%%%%%%%%%%%%%%%%%%%%%

\section{VdW potential near an interface}
\label{sec_twolayer}

Let us first apply the theory to the simplest case of a single
interface between two homogeneous semi-infinite magnetoelectric half
spaces, where we first concentrate on the position-dependent part of
the potential responsible for the vdW force and then interpret the
(layer-dependent) constant part.
 
%%%%%%%%%%%%%%%%%%%%%%%%%%%%%%%%%%%%%%%%%%%%%%%%%%%%%%%%%%%%%%%%%%%%%%

\subsection{Position-dependent part}

For a planar magnetoelectric two-layer system with the guest atom
placed in, say, layer $2$, substitution of $r^\sigma_{n+}=0$ and 
$r^\sigma_{n-}$ in accordance with Eqs.~(\ref{equ9b}) in
Eq.~(\ref{Umiddle}) leads to the following expression for
the position-dependent part of the vdW potential
\begin{multline}
\label{U2_2layer}
U_2(z_A)= \frac{\hbar \mu_0}{8\pi^2}\int _0^{\infty} \mathrm{d}\xi\, 
\xi^2  \alpha \left(\frac{3\varepsilon_2}{2\varepsilon_2+1}\right)^2 
\\
\times \mu_2 \int _0^{\infty} \mathrm{d}q\, \frac{q}{\beta_2}
\left[\frac{\mu_1 \beta_2-\mu_2 \beta_1}{\mu_1
\beta_2+\mu_2 \beta_1}\right.\\
-\left.\frac{\varepsilon_1 \beta_2-\varepsilon_2 \beta_1}
{\varepsilon_1 \beta_2 + \varepsilon_2
\beta_1}\left(1+2\frac{q^2c^2}{\varepsilon_2\mu_2
\xi^2}\right)\right]
e^{-2 \beta_2 z_A},
\end{multline}
which differs from its atom-in-free-space counterpart by the 
local-field correction factor 
$\left(\frac{3\varepsilon_2}{2\varepsilon_2+1}\right)^2$
and by $\varepsilon_2\neq 1$ and $\mu_2\neq 1$ \cite{Kampf2005}
%%%%%%%%%%%%%%%%%%%%%%%%%%%%%%%%%%%%%%%%%

\subsubsection{Retarded limit}
\label{retarded_limit}

It is instructive to study the potential in the retarded limit
\begin{equation}
z_A>>\frac{c}{\omega_A^-}\ \mathrm{and}\ 
z_A>>\frac{c}{\omega_M^-},
\end{equation} 
where $\omega_A^-$ and $\omega_M^-$ being the minima of all relevant
atomic transition and medium resonance frequencies, respectively.
Introducing a new integration variable \mbox{$v=c \beta_2 /\xi$} and
replacing \mbox{$\alpha(i\xi)\simeq\alpha(0)$},
\mbox{$\varepsilon_{1,2}(i\xi)\simeq\varepsilon_{1,2}(0)$}, and
\mbox{$\mu_{1,2}(i\xi)\simeq\mu_{1,2}(0)$}, the integration over $\xi$
can be performed and we obtain for the potential
\begin{equation}
\label{Ublong}
U_2(z_A)=\frac{C_4}{z_A^4},
\end{equation}
where
\begin{multline}
\label{C1}
C_4\!=\!\frac{3\hbar c}{64\varepsilon_0\pi^2}\alpha(0)\!
\left(\frac{3\varepsilon_2(0)}{2\varepsilon_2(0)+1}\right)^2\!\mu_2(0)
\!  
\int _{\sqrt{\varepsilon_2(0)\mu_2(0)}}^{\infty}\mathrm{d}v\, \\
\times\frac{1}{v^4} 
\left[\frac{\mu_1(0) v-\mu_2(0)\sqrt{v^2-\varepsilon_2(0)
\mu_2(0)+\varepsilon_1(0)\mu_1(0)}}{\mu_1(0)
v+\mu_2(0)\sqrt{v^2-\varepsilon_2(0)\mu_2(0)+\varepsilon_1(0)\mu_1(0)
}} \right.
\\
+\left. \frac{\varepsilon_1(0) v-\varepsilon_2(0)\sqrt{v^2-
\varepsilon_2(0)
\mu_2(0)+\varepsilon_1(0)\mu_1(0)}}{\varepsilon_1(0)
v+\varepsilon_2(0)\sqrt{v^2-\varepsilon_2(0)\mu_2(0)+\varepsilon_1(0)
\mu_1(0)}}
\right. \\
\left.\times(1-2\frac{v^2}{\varepsilon_2(0)\mu_2(0)})\right].
\end{multline}
It can be proven that 
$\partial C_4/\partial \varepsilon_1(0) <0$,
$\partial C_4/\partial \mu_1(0) >0$,  and
$\partial C_4/\partial \mu_2(0) <0$.

To deduce some physics from Eq.~(\ref{C1}), we consider some limiting 
cases. Assuming that the atom is located in free space 
$\mu_2(0)=\varepsilon_2(0)=1$, it can be shown that,
for a purely electric halfspace $1$, 
\begin{equation}
\label{C1.1}
C_4 [\mu_1(0)=1, \mu_2(0)=1,\varepsilon_2(0)=1] < 0
\end{equation}
and, for a purely magnetic halfspace $1$ with $\mu_1(0)>1$,
\begin{equation}
\label{C1.2}
C_4 [\varepsilon_1(0)=1,\mu_2(0)=1,\varepsilon_2(0)=1] > 0. 
\end{equation}
The inequality (\ref{C1.1}) means the atom is attracted toward the
electric halfspace --- the case which is commonly treated in earlier
literature. On the other hand, the positivity of  $C_4$ in Eq.
(\ref{C1.2}) means the atom is repelled from the magnetic half space.
This, coupled with the signs of the derivatives given below
Eq.~(\ref{C1}), imply that electric properties tend to make the
potential attractive while magnetic ones tend to make the potential
repulsive.

Now if the atom is embedded in a material halfspace, while 
the opposite halfspace is empty $\mu_1(0)=\varepsilon_1(0)=1$,
it can be shown that, for a purely electric material, 
\begin{equation}
\label{C1.3}
C_4 [\mu_1(0)=1,\varepsilon_1(0)=1,\mu_2(0)=1] > 0,
\end{equation}
the atom is repulsed from the interface, while for a purely magnetic
material with $\mu_2(0)>1$,
\begin{equation}
\label{C1.4}
C_4 [\mu_1(0)=1,\varepsilon_1(0)=1, \varepsilon_2(0)=1] < 0, 
\end{equation}
the atom is attracted towards the interface. Since $\partial
C_4/\partial \mu_2(0) <0$, if one starts from a purely electric
material which is accompanied by a repulsive potential, then enhances
the magnetic properties of the material by increasing $\mu_2(0)$, one
would obtain with an attractive potential in accordance with
Eq.~(\ref{C1.4}).

Another particular case is when the magnetodielectric contrast between
the contacting media is small
\begin{align}
    & \varepsilon_1(0)=\varepsilon_2(0)+\chi(0),\\
    & \mu_1(0)=\mu_2(0)+\zeta(0)
\end{align}
[$\chi(0)\ll \varepsilon_2(0)$, $\zeta(0)\ll \mu_2(0)$], one can
further treat the integrals analytically by keeping only terms linear
in $\chi$ and $\zeta$,
\begin{align} 
& \frac{\mu_1(0) v-\mu_2(0)\sqrt{v^2-\varepsilon_2(0)
\mu_2(0)+\varepsilon_1(0)\mu_1(0)}}{\mu_1(0)
v+\mu_2(0)\sqrt{v^2-\varepsilon_2(0)\mu_2(0)+\varepsilon_1(0)\mu_1(0)}}
\nonumber \\
& \hspace{1cm} 
  \simeq
\left(\frac{1}{2\mu_2(0)}-\frac{\varepsilon_2(0)}{4v^2}\right)\zeta(0)
-  \frac{\mu_2(0)}{4v^2}\chi(0),
\\ 
& \frac{\varepsilon_1(0) v-\varepsilon_2(0)\sqrt{v^2-\varepsilon_2(0)
\mu_2(0)+\varepsilon_1(0)\mu_1(0)}}{\varepsilon_1(0)
v+\varepsilon_2(0)\sqrt{v^2-\varepsilon_2(0)\mu_2(0)+\varepsilon_1(0)
\mu_1(0)}}
\nonumber\\
& \hspace{1cm}
\simeq -\frac{\varepsilon_2(0)}{4v^2}\zeta(0)
+\left(\frac{1}{2\varepsilon_2(0)}-
\frac{\mu_2(0)}{4v^2}\right)\chi(0).
\end{align}
The $v$-integration is straightforward and we arrive at
\begin{multline}
\label{C1approx}
    C_4=\frac{9\hbar c}{640\pi^2\varepsilon_0}\alpha(0)
   \frac{-23\mu_2(0)\chi(0)+7\varepsilon_2(0)\zeta(0)}{
   \sqrt{\varepsilon_2(0)\mu_2(0)}\mu_2(0)[2\varepsilon_2(0)+1]^2}.
\end{multline}
This result generalises the one obtained in Ref.~\cite{Kampf2005} to 
the case of an atom embedded in a medium, with local-field correction
included. It is richer in content than its atom-in-free-space 
counterpart. For example, when $\chi(0)=\zeta(0)$ and the atom in
free space $\varepsilon_2(0)=\mu_2(0)=1$, the potential is attractive,
whereas when the atom is in a medium of
$\varepsilon_2(0)/\mu_2(0)>23/7$, a repulsive potential can be
realized.

%%%%%%%%%%%%%%%%%%%%%%%%%%%%%%%%%%%%%%%%%%%
\subsubsection{Non-retarded limit}

The non-retarded limit corresponds to atom--surface distances $z_A$ 
small compared with the typical wavelengths of the medium and the
atomic system, 
\begin{align}
\label{short1}
z_A<<&\frac{c}{\omega_A^+[n_1(0)+n_2(0)]}\quad\mathrm{and/or}\\
\label{short2}
z_A<<&\frac{c}{\omega_M^+ [n_1(0)+n_2(0)]}
\end{align} 
[$n_{1,2}(0)\!=\!\sqrt{\varepsilon_{1,2}(0) \mu_{1,2}(0)}$; 
\mbox{$\omega_A^+$} and \mbox{$\omega_M^+$}, maxima of the relevant
atomic transition and medium resonance frequencies]. The conditions
(\ref{short1}) and (\ref{short2}) imply
\begin{align}
\label{45}
&\frac{\xi z_A}{c}\sqrt{\left|\varepsilon_1\mu_1-\varepsilon_2
\mu_2\right|}<<1,\\
\label{46}
&\frac{\xi z_A}{c}\sqrt{\varepsilon_2\mu_2}<<1,
\end{align}
where we have used that the $\xi$-integration is practically limited
to a region where $\xi\lesssim \omega_{A,M}^+$. We substitute
\begin{align} 
    &q=\sqrt{\beta_2^2-\varepsilon_2\mu_2 \xi^2/c^2},\\
    & q \mathrm{d}q= \beta_2\mathrm{d}\beta_2,\\
    & \beta_1= \sqrt{\xi^2/c^2(\varepsilon_1\mu_1-\varepsilon_2\mu_2)+
                    \beta_2^2}
\end{align}
in Eq.~(\ref{U2_2layer}) and, on recalling Eq.~(\ref{45}),
perform a leading-order Taylor expansion in
$\xi^2/(c^2\beta_2^2)(\varepsilon_1\mu_1-\varepsilon_2\mu_2)$.
After carrying out the $\beta_2$-integral, we arrive at
\begin{equation}
\label{Ubshort}
U_2(z_A)=-\frac{C_3}{z_A^3}+\frac{C_1}{z_A},
\end{equation}
where
\begin{equation}
\label{C3}
C_3=\frac{\hbar}{16\pi^2\varepsilon_0}\int _0^{\infty}\mathrm{d}\xi
\alpha\frac{9\varepsilon_2}{\left(2\varepsilon_2+1\right)^2}
\frac{\varepsilon_1-\varepsilon_2}{\varepsilon_1+\varepsilon_2},
\end{equation}
\begin{align}
\label{C2}
&C_1  = \frac{\hbar\mu_0}{16\pi^2}\int _0^{\infty} \mathrm{d}\xi \xi^2
\alpha
\mu_2 \left(\frac{3\varepsilon_2}{2\varepsilon_2+1}\right)^2 
\nonumber\\ 
& \times\left[\frac{\mu_1-\mu_2}{\mu_1+\mu_2}+\frac{
\varepsilon_1-\varepsilon_2}{\varepsilon_1+\varepsilon_2}
+\frac{2\varepsilon_1(\varepsilon_1\mu_1-\varepsilon_2\mu_2)}
{\mu_2(\varepsilon_1+
\varepsilon_2)^2}
\right],
\end{align}
and Eq.~(\ref{46}) has been used to set
$\exp(-2\sqrt{\varepsilon_2\mu_2}\xi z_A/c)$ equal to one. By putting 
$\varepsilon_2=\mu_2=1$, Eqs.~(\ref{C3}) and (\ref{C2}) reduce to
those for an atom in free space \cite{Kampf2005}.

As can be seen from Eqs.~(\ref{Ubshort})--(\ref{C3}), one can
distinguish two regimes having different power laws. The first regime
is where the two contacting media have unequal electric properties.
Then the first term in Eq.~(\ref{Ubshort}) dominates and the power law
is $z_A^{-3}$. The atom is pulled toward (repelled from) the interface
if the medium it is located in has stronger (weaker) electric
properties than that on the other side of the interface. 

In the case of equal electric properties $C_3=0$ and
$U_2(z_A)=C_1/z_A$ with 
\begin{align}
\label{C2a}
&C_1  = \frac{\hbar\mu_0}{16\pi^2}\int _0^{\infty} \mathrm{d}\xi \xi^2
\alpha
\left(\frac{3\varepsilon_2}{2\varepsilon_2+1}\right)^2 
(\mu_1-\mu_2)
\nonumber\\ 
& \hspace{2cm}\times\left(\frac{\mu_2}{\mu_1+\mu_2} + \frac{1}{2} 
\right).
\end{align}
The atom experiences a force which points away from (toward) the
interface if the magnetic properties of the medium the atom is
situated in are weaker (stronger) than those of the medium on the
other side of the interface. 

Note that the dependence of the directions of the forces on the
difference in strength of the medium responses is opposite in the two
cases of dominantly electric and purely magnetic media. In both cases
the strength of the force increases with increasing difference between
the electric and magnetic parameters of the contacting media. These
results are consistent with earlier ones
\cite{Langbein,Israelachvili1974}. A more detailed comparison of our
results with those presented in Ref.~\cite{Israelachvili1974} will be
made in the last section. Since the coefficient $C_3$ of the
leading-order term depends on electric properties only, the influence
of electric properties on the behaviour of the potential tends to
dominate that of the magnetic properties, except for the case when the
electric properties of the neighboring media are similar. At more
moderate distances, competing effects of electric and magnetic
properties may create potential walls or wells, as is also evident
from the numerical results below.

%%%%%%%%%%%%%%%%%%%%%%%%%%%%%%%%%%%%%%%%%%%%%%%%%%%%%%

\subsubsection{Numerical results}

To study the local-field corrected potential at intermediate distances
and to elucidate the combined influence of electric and magnetic
properties of the media, we calculate the position dependent part
$U_2(z_A)$ in accordance with Eq.~(\ref{U2_2layer}) numerically. A
two-level atom of transition frequency $\omega_{10}$ is assumed and 
material electric and magnetic properties are described using
single-resonance Drude-Lorentz-type permittivities and permeabilities
\begin{align}
   &\varepsilon_i(\omega) = 
          1+ \frac{\omega_{Pei}^2}{\omega_{Tei}^2-\omega^2-
          i\omega\gamma_{ei}},
\\
   & \mu_i(\omega) = 1 +\frac{\omega_{Pmi}^2}{\omega_{Tmi}^2
            -\omega^2-i\omega\gamma_{mi}}, 
\quad i=1,2.
\end{align}

\begin{figure}[!t!]
\begin{center}
\includegraphics[width=\linewidth]{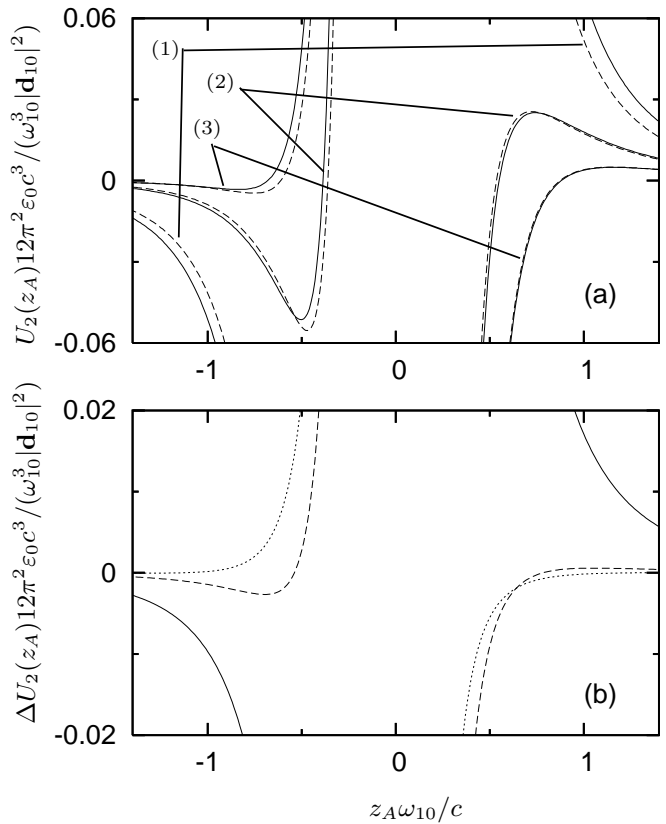}
\end{center}
\caption{(a) 
Position-dependent part of the vdW potential experienced by a
ground-state two-level atom in a magnetoelectric two-layer system as a
function of atom--surface distance for fixed $\varepsilon_1$, $\mu_1$,
$\mu_2$, and for $\omega_{Pe2}/\omega_{10}=1$ (1), $0.4$ (2), and
$0.2$ (3). Solid lines denote the potentials with the local-field
correction, while dashed lines represent those without. Other
parameters are
$\omega_{Te1}/\omega_{10}=\omega_{Te2}/\omega_{10}=1.03$,
$\omega_{Pe1}/\omega_{10}=0.75$,
$\omega_{Tm1}/\omega_{10}=\omega_{Tm2}/\omega_{10}=1$, 
$\omega_{Pm1}/\omega_{10}=2.3$,
$\omega_{Pm2}/\omega_{10}=0.4$, 
$\gamma_{m1,2}/\omega_{10}=\gamma_{e1,2}/\omega_{10}=0.001$,
and the cavity radius is $R_\mathrm{c}\omega_{10}/c=0.01$.
(b) Difference between local-field corrected and uncorrected
(position-dependent) vdW potential $\Delta U_2$ versus atom--surface
distance where the solid, dashed, and dotted lines refer to the curves
(1), (2), and (3), respectively. 
} 
\label{za1}
\end{figure}

Figure~\ref{za1}~(a) illustrates the atom--surface-distance dependence
of the $U_2(z_A)$ potential. The interface is at $z_A=0$, the medium 1
is on the left, while the medium 2 is on the right. The parameters
$\varepsilon_1$, $\mu_1$, $\mu_2$ are fixed whereas $\varepsilon_2$
takes on three different values of the plasma frequency
$\omega_{Pe2}$. In all three cases $\varepsilon_1$ and $\varepsilon_2$
have same (transverse) resonance frequencies $\omega_{Te}$ and damping
constants $\gamma_e$, but different plasma frequencies (i.e.,
$\varepsilon_1\neq \varepsilon_2$) and the $z_A^{-1}$ term in Eq.
(\ref{Ubshort}) is negligibly small. In case (1) $\varepsilon_2 >
\varepsilon_1$, it can be seen from the figure that the potential at
very short atom--surface distances is repulsive in medium 2 and 
attractive in medium 1, in consistency with the analytical result
(\ref{Ubshort}) and (\ref{C3}). Similarly, in cases (2) and (3)
$\varepsilon_2 < \varepsilon_1$, the potential is attractive in medium
2 while repulsive in medium 1. As the atom--surface distance
increases, the second term $\frac{C_1} {z_A}$ in the potential
(\ref{Ubshort}) gradually comes into play and if the magnetic
properties are strong enough, may switch the sign of the potential and
create potential walls or wells in the process, as is clearly visible
in case (2).

%%%%%%%%%%%%%%%  F I G U R E %%%%%%%%%%%%%%%%%%%%%%
%%%%%%%%%%%%%%%%%%%%%%%%%%%%%%%%%%%%%%%%%%%%%%%%%%%%%%%%%%

To show how the net effect of the local-field correction depends on
the distance and on the properties of the media surrounding the guest
atom, we plot the uncorrected potential by dashed lines in
Fig.~\ref{za1}~(a), and the difference between the corrected and
uncorrected results in Fig.~\ref{za1}~(b). The ratio between the
corrected and uncorrected results is not always a good measure of the
difference between the two because one of them can vanish. The
local-field correction factor $[3\varepsilon_j/(2\varepsilon_j+1)]^2$
is positive, larger than one, and increases with $\varepsilon_j$ 
[$j$ indicative of the layer containing the guest atom]. It approaches
the maximum value of $9/4$ as $\varepsilon_j \rightarrow \infty$. Note
that a larger-than-unity local-field correction factor does not
necessarily lead to an enhancement of the potential because the
uncorrected factor in the integrand can change sign as $\xi$ varies.
The local-field correction has a clear-cut effect of increasing or
decreasing the potential only when the uncorrected integrand is purely
repulsive or attractive, for which cases (1) and (3) can serve as
examples. In the middle case (2), the two curves with and without
local-field correction cross, that is, there exists an atom--surface
distance at which the effect of the local-field correction is
canceled out due to the $\xi$-integration. In addition, one can
notice that the local-field correction leads a small shift of the
position of the peak. Figure~\ref{za1}~(b) which shows the difference
between the local-field corrected and uncorrected potentials, reveals
quite significant corrections of up to 30\% of the uncorrected
values. 

%%%%%%%%%%%%%%%  F I G U R E %%%%%%%%%%%%%%%%%%%%%%
\begin{figure}[!t!]
\begin{center}
\includegraphics[width=\linewidth]{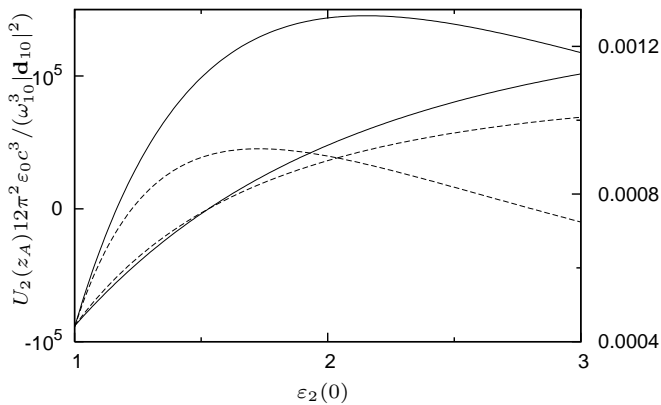}
\end{center}
\caption{ 
Position-dependent part of the vdW potential as a function of the
static permittivity $\varepsilon_2(0)$ (more specifically
$\omega_{Pe2}/\omega_{10}$) for two values of the atom--surface
distance $z_A\omega_{10}/c=0.01$ (scale to the left) and
$z_A\omega_{10}/c=3$ (scale to the right). Solid lines are with the
local-field correction while dashed lines are without one. Other
parameters are the same as in Fig.~\ref{za1}. Note the arrows which
indicate the ordinate scale to be used.}
\label{ope}
\end{figure}
%%%%%%%%%%%%%%%  F I G U R E %%%%%%%%%%%%%%%%%%%%%%
The behaviour of the local-field corrected vdW potential with respect
to the static permittivity of the medium the atom is embedded in is
shown in Fig.~\ref{ope} for two different values of the atom--surface
distance. Within the scale of the figure, the curves for the larger
distance from the interface peak at certain values of
$\varepsilon_2(0)$. The positions of the peaks are different due to
the effects of the local field. As $\omega_{Pe2}/\omega_{10}$ and as a
consequence $\varepsilon_2(0)$ increases, an inspection of the figure
reveals that the ratio between the corrected and uncorrected curves
tends to the static value of the local-field correction factor 
$[3\varepsilon_2(0)/(2\varepsilon_2(0)+1)]^2$ (which lies between $1$
and $9/4$), in agreement with the analytical analysis given in
Sec.~\ref{retarded_limit}. For the smaller value of the atom--surface
distance $z_A\omega_{10}/c=0.01$, a crossing point between the
corrected and uncorrected curves is observed, where the local-field
correction produces no net change. 

%%%%%%%%%%%%%%%%%%%%%%%%%%%%%%%%%%%%%%%%%%%%%%%
\subsection{
Layer-dependent constant part
}

%%%%%%%%%%%%%%%  F I G U R E %%%%%%%%%%%%%%%%%%%%%%
\begin{figure}[!t!]
\begin{center}
\includegraphics[width=\linewidth]{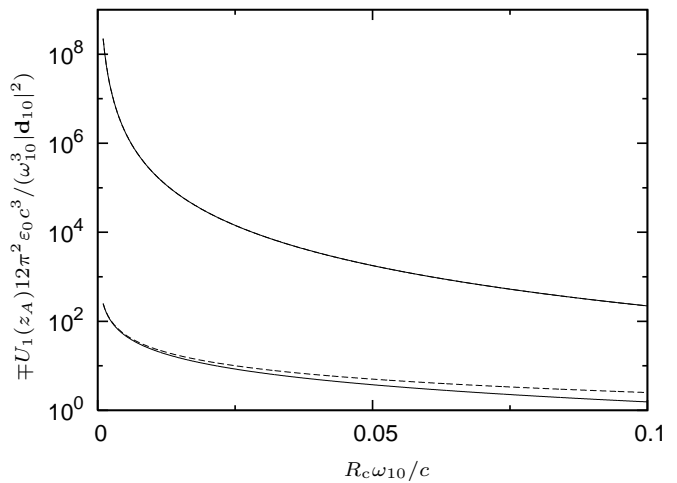}
\end{center}
\caption{
The exact layer-dependent constant part of the potential (solid line),
Eq.~(\ref{U1}), and approximate results (dashed line),
Eq.~(\ref{U1approx}), are shown as functions of the real-cavity
radius. The upper pair of curves shows $-U_1(z_A) 12 \pi^2
\varepsilon_0 c^3/(\omega^3_{10} | {\bf d}_{10}|^2)$ (the sign has
been reversed so that a  logarithmic scale can be used) for a pure
electric material with $\omega_{Pe2}/\omega_{10}=0.4$, while the lower
pair of curves shows $U_1(z_A) 12 \pi^2 \varepsilon_0
c^3/(\omega^3_{10} |
{\bf d}_{10}|^2)$ for a pure magnetic material with
$\omega_{Pm2}/\omega_{10}=0.4$. All other parameters are the same as
in Fig.~\ref{za1}. The radius of the cavity
$R_\mathrm{c}\omega_{10}/c$ starts from $0.001$.}
\label{Rc}
\end{figure}
%%%%%%%%%%%%%%%%%%%%%%%%%%%%%%%%%%%%%%%%%%%%%%%

Fig.~\ref{Rc} shows the dependence of the constant part $U_1$ of the
potential on the real-cavity radius $R_\mathrm{c}$. To gain some
insight, we consider the two limiting cases of a purely electric and a
purely magnetic material. According to the analytic result
(\ref{U1approx}), in the first case the leading term is proportional
to $[R_\mathrm{c}\omega_{10}/c]^{-3}$, while in the second case the
leading term is proportional to $[R_\mathrm{c}\omega_{10}/c]^{-1}$.
Thus for small enough $R_\mathrm{c}\omega_{10}/c$, $|U_1|$ for a pure
electric material is generally larger than that for a pure magnetic
material - a fact that is confirmed by the figure. It can be seen that
$U_1<0$ in the first case and $U_1>0$ in the second case, again in
agreement with Eq.~(\ref{U1approx}). The figure also indicates that
the magnitude of $U_1$, which is entirely due to the local-field
correction, decreases with an increasing real-cavity radius, that is,
the effects of the local field becomes weaker as the medium becomes
more dilute.

For comparison, we also plot the potential $U_1$ according to the 
approximate result (\ref{U1approx}) (dashed curves). It can be seen 
that for the parameters used in the figure, the approximate result
reproduces quite well the exact one, especially in the case of a pure
electric material. The agreement in the case of a pure magnetic
material is good for very small $R_\mathrm{c}\omega_{10}/c$, but
worsens as $R_\mathrm{c}\omega_{10}/c$ increases.
 
%%%%%%%%%%%%%%%%%%%%%%%%%%%%%%%%%%%%%%%%%%%%%%%%%%%%%%%%%

\subsection{Total vdW potential and its value at the interface}

\begin{figure}[!t!]
\begin{center}
\includegraphics[width=\linewidth]{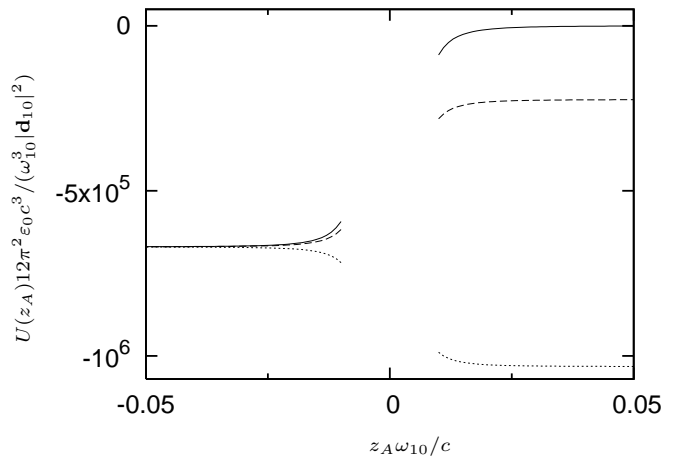}
\end{center}
\caption{Local-field corrected total vdW potential of a ground-state 
two-level atom in a magnetodielectric two-layer system as a function
of the atom--surface distance. Different curves are for different
(equal) electric and magnetic coupling strengths of the medium 2:
$\omega_{Pm2}/\omega_{10}=\omega_{Pe2}/\omega_{10}=0$ (solid line),
$0.4$ (dashed line) and $1$ (dotted line). Other parameters are the
same as in Fig.~\ref{za1}. 
} 
\label{UC}
\end{figure}

We have separately investigated the position-depen\-dent part
$U_2(z_A)$ of the potential, which determines the force acting on the
atom, and the constant, layer-dependent part $U_1$, which is related
to the local-field correction. For problems such as the transfer of an
atom or a small particle through an interface, it is of relevance to
evaluate the potential right at the interface. For this purpose, the
total value of the potential is needed. In Fig.~\ref{UC}, we have
calculated $U_1 + U_2(z_A)$ on both sides of the interface with the 
medium 1 fixed while the medium 2 varying from vacuum to a more dense
medium with balanced electric and magnetic properties. The case
represented by dashed line is nothing else rather than case (2) in
Fig.~\ref{za1}. Only very short distances
$\sqrt{|\varepsilon_i\mu_i|}z_A\omega_{10}/c\ll 1$ are presented.
Since $\varepsilon_1 \neq \varepsilon_2$ in general, the position
dependence of the potential at short distances is mostly determined by
the $\frac{C_3}{z_A^3}$ term which contains $\varepsilon_1 -
\varepsilon_2$ in the integrand. Numerical results in the figure are
consistent with this and show that the potential is attractive
(repulsive) if the atom is located in a medium which is electrically
more dilute (dense) than that in the opposite side of the interface.
Additional structures in $U_2$ like potential wells or walls are
typically overwhelmed by the magnitude of $U_1$. Baring the visual
suppression of the potential wells or walls in $U_2$ due to large
magnitude of $U_1$, a full potential is more straightforward than
$U_2$ alone in predicting the movement of an atom across a surface.
Take, for example, case (2) in Fig.~\ref{za1}. An atom located in
layer 2 close to the surface will be attracted to it, and if the atom
can cross the interface, it will be pushed further away from the
surface into layer 1. Fig.~\ref{UC} (dashed line) provides us with
some additional information. It shows explicitly that the total
potential in layer 2 is higher that that in layer 1, by giving the
difference between the potentials in the two layers; and it may also
help to estimate the amount of energy required for the atom to
penetrate into layer 1.

%%%%%%%%%%%%%%%  F I G U R E %%%%%%%%%%%%%%%%%%%%%%
%%%%%%%%%%%%%%%%%%%%%%%%%%%%%%%%%%%%%%%%%%%%%%%%%%%%%%%%%%%%%%%%%%%%%

In Fig.~\ref{UC}, we have not displayed the results for distances
$|z_A| < R_\mathrm{c}\sqrt{|\varepsilon \mu|}$ where the real-cavity
model can no longer be applied. This gives rise to a gap between the
potentials on the two sides of the interface. To extend our theory so
that it can be employed to estimate the potential right at the
interface, we suggest that
\begin{multline}
\label{interface}
U(z_A=0)=\frac{1}{2}\left[U(R_\mathrm{c})+U(-R_\mathrm{c})\right]\\
=-\frac{\hbar}{32\pi^2\varepsilon_0R_\mathrm{c}^3}\int _0^{\infty}
\mathrm{d}\xi\,
\alpha\Biggl\{12\left(\frac{\varepsilon_1\!-\!1}{2\varepsilon_1\!+\!1}
+\frac{\varepsilon_2\!-\!1}{2\varepsilon_2\!+\!1}\right)\\
-\,
\frac{\varepsilon_1\!-\!\varepsilon_2}
{\varepsilon_1\!+\!\varepsilon_2}\left[
\frac{1}{\varepsilon_1}
\left(\frac{3\varepsilon_1}{2\varepsilon_1\!+\!1}\right)^2
-\frac{1}{\varepsilon_2}
\left(\frac{3\varepsilon_2}{2\varepsilon_2\!+\!1}\right)^2\right]
\Biggr\}
\end{multline}
[recall Eqs.~(\ref{U1approx}), (\ref{Ubshort}), and (\ref{C3})].
Visually, this means first plotting the potential as a function of the
atomic position up to distances $|z_A|=R_{\rm c}$ ($R_{\rm c}$ being
the radius of the cavity in the real-cavity model), then connecting
the two loose ends on the two sides of the interface to find the value
of the potential at $z_A=0$.
Our result is remarkably similar to what has been obtained by
calculating the on-surface potential of a molecule of finite size $s$
\cite{NinhamBook},
\begin{multline}
\label{Ninhamresult}
U(z_A=0)=\frac{\hbar}{2\pi^{5/2}\varepsilon_0s^3}\int
_0^{\infty}
\mathrm{d}\xi\,
\alpha\biggl[\frac{1}{2}\left(\frac{1}{\varepsilon_1}
+\frac{1}{\varepsilon_2}\right)\\
+\,\frac{1}{3}\frac{\varepsilon_1\!-\!\varepsilon_2}
{\varepsilon_1\!+\!\varepsilon_2}\left(
\frac{1}{\varepsilon_1}
-\frac{1}{\varepsilon_2}\right)
\biggr].
\end{multline}
The second terms in Eqs.~(\ref{interface}) and (\ref{Ninhamresult}),
which represent the interface contribution to the potential, agree
when setting $s=(\sqrt[3]{16/3}/\pi^{-1/6})R_\mathrm{c}\approx
1.4R_\mathrm{c}$ and neglecting the local-field correction in
Eq.~(\ref{interface}) which was not considered in
Ref.~\cite{NinhamBook}. The first terms, which represent bulk
contributions from the two interfacing media, differ in the two
approaches, where Eq.~(\ref{Ninhamresult}) still contains self-energy
contributions which do not vanish in the vacuum case $\varepsilon_i=1$
while Eq.~(\ref{interface}) does not. Our result~(\ref{interface})
thus represents an improvement of the previous
one~(\ref{Ninhamresult}) in that local-field corrections are taken
into account and self-energy contributions have consistently been
removed.

%%%%%%%%%%%%%%%%%%%%%%%%%%%%%%%%%%%%%%%%%%%%%%%%%

\section{Three-layer system}
\label{subsec_three}

A system consisting of an atom in a three-layer planar structure can
serve as a prototype for the problem of a small particle near or
inside a membrane \cite{Israelachvili1974}. The translationally
invariant part and position-dependent part of the vdW potential can
again be determined in accordance with Eqs.~(\ref{U1}) and
(\ref{Umiddle}), respectively. If the atom is in one of the two outer
layers, Eq. (\ref{Umiddle}) simplifies greatly via either
$r^\sigma_{j+}=0$ or $r^\sigma_{j-}=0$. If the atom is in the middle
layer, the overall form of the potential $U_2(z_A)$ remains as in
Eq.~(\ref{Umiddle}) and there exists a term which contains the product
$r^\sigma_{j-} r^\sigma_{j+}$ and is position-independent. This
position-independent term in $U_2(z_A)$ is irrelevant when it is the
force which is concerned, but has to be kept, together with $U_1$, if
the potential at the surfaces is of interest. Since the formulas are
complicated, we resort to numerical computation.

%%%%%%%%%%%%%%%  F I G U R E %%%%%%%%%%%%%%%%%%%%%%
\begin{figure}[!t!]
\begin{center}
\includegraphics[width=\linewidth]{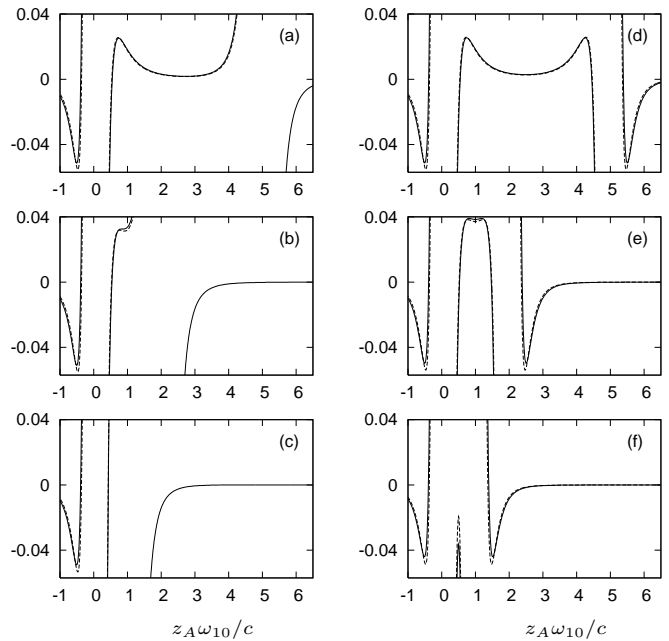}
\end{center}
\caption{
 Position-dependent part of the vdW potential as a function of 
 $z_A\omega_{10}/c$ for three different thicknesses of the middle
 layer
 $d_2\omega_{10}/c=$ 5 [(a) and (d)], 2 [(b) and (e)], and 1 [(c) and
 (f)].
 For the left column, the configuration is asymmetric with
 $\omega_{Te1}/\omega_{10}=1.03$, $\omega_{Pe1}/\omega_{10}=0.75$,
 $\omega_{Tm1}/\omega_{10}=1$, $\omega_{Pm1}/\omega_{10}=2.3$,
 $\omega_{Tm2}/\omega_{10}=1$, $\omega_{Pm2}/\omega_{10}=0.4$,
 $\omega_{Te2}/\omega_{10}=1.03$, $\omega_{Pe2}/\omega_{10}=0.4$,
 $\gamma_{m2,1}=\gamma_{e2,1}=0.001\omega_{10}$, 
 $\varepsilon_3=\mu_3=1$, and
 $R_\mathrm{c}\omega_{10}/c=0.01$.
 For the right column, the configuration is symmetric with the vacuum 
 in layer 3 being replaced by a medium of the same characteristics as
 those of layer 1. The curves without the local-field correction are
 shown by dashed lines.
 }
 \label{a3}
\end{figure}%
%%%%%%%%%%%%%%%%%%%%%%%%%%%%%%%%%%%%%%%%%%%%%%%%%%%%%%%%%

Figure~\ref{a3} shows the behaviour of the vdW potential for an atom
placed in an asymmetric (left column) and a symmetric (right column) 
three-layer magnetodielectric structure for different thicknesses of
the middle layer. Note that the parameters for layer 1 are the same as
those for layer 1 in Fig.~\ref{za1} and the parameters for layer 2 are
the same as those for layer 2, case (2), in Fig.~\ref{za1}. That is,
the interface 1-2 here is the same as the interface 1-2 in
Fig.~\ref{za1}(2). In the asymmetric configuration, layer 3 is vacuum
while in the symmetric configuration, layer 3 and 1 have the same
characteristics.

Let us first analyse the asymmetric configuration and see how the
presence of a third layer 3 on the right, which is vacuum, affects the
behaviour of the potential near the 1-2 interface. In case (a) where
the middle-layer thickness $d_2\omega_{10}/c=5$ is the largest among
three cases, the behaviour of the potential $U_2(z_A)$ in the boundary
regions is similar to that in the two-layer systems. Near the 1-2
interface, we find a potential well in layer 1 and a potential wall in
layer 2, just as in Fig.~\ref{za1}(2). Near the 2-3 interface the
potential is repulsive in the more dense medium 2 and attractive
in the more dilute medium 3. A new feature appears around the center
of the middle layer where a finite potential wall on the left and an
attractive one towards the right interface combine to a potential
well. Clearly, if an atom at rest is put in the well, it will remain
there. When the thickness of the middle layer $2$ is reduced, the well
becomes more shallow and eventually disappears, as is visible in
Figs.~\ref{a3}(b) and (c). For the parameters used in the figure, the
potential well in the middle layer occurs when $d_2\omega_{10}/c \gg
1$ and is overwhelmed when $d_2\omega_{10}/c \sim 1$. The curves for
the potential can help one to predict the movement of an atom located
near or inside a membrane (layer 2). For example, let an atom be
initially located in layer 3 (vacuum) of case (a). First, it will be
attracted to the 2-3 interface. If it can be transported through the
interface and gather enough momentum going down the slope, it can pass
the finite potential wall and is then attracted to the 1-2 interface.
After this interface, and maybe some oscillations, it will be
suspended in the potential well near the surface.

We now turn to the symmetric configuration (right column of
Fig.~\ref{a3}) where the middle layer 2 is sandwiched between two
identical layers 1 and 3. The behaviour of the potential on the right
is just a mirror image of that on the left with the mirror plane being
the one parallel to the surface and passing through the centre of the
middle layer. When the thickness of layer $2$ is largest
$d_2\omega_{10}/c =5$ [case (d)], we see in layer 2 a combination of 
two potential walls as found in Fig.~\ref{za1}, with a well in the
middle. Thus even as the middle layer is in general less dense than
the two surrounding layers in both electric and magnetic aspects,
there exists a possibility that an atom initially located in the
potential well remains there. With decreasing thickness $d_2$, the
bottom of the well rises and the two walls eventually merge into one
[cases (e) and (f)]. Clearly, in the cases (e) and (f), any atom
initially situated in the middle layer will be transported to the
neighboring layers. Numerical computation also shows that the
magnitude of the position-independent term in $U_2(z_A)$, i.e., the
last term in Eq.~(\ref{Umiddle}), is negligible compared to the
position-independent terms, due to the small exponential factor.

In Fig.~\ref{a3}, the uncorrected potentials are  plotted by dashed
lines. In the two outer layers, the effects of the local field almost
remain the same as in the two-layer configuration, which can be
explained by that the presence of a third layer is screened by the
middle layer. For the middle layer, we have found that typically, the
local-field correction is most significant around the centre of the
layer, as is most manifest in case (f) where a correction 
[$(U_{2{\rm corrected}}-U_{2{\rm uncorrected}})/U_{2{\rm
uncorrected}}$] of more than 80\%  is observed. A variation of the
middle layer thickness does not affect much the strength of the local
field correction near the surfaces. This can be understood as
resulting from the fact that when an atom is located very close to a
surface, it will tend to see only the nearest neighboring layer.

If one wish to know the potentials right on the surfaces, one would
have to evaluate the full potential $U_1+U_2(z_A)$ for atom--surface
distances large enough such that the macroscopic theory applies, then
use it as an input to the procedure proposed in the previous section.
Since the position-independent part $U_1$ does not depend on the layer
thicknesses, at each surface in a more-than-two-layer system, the
results will be closely analogous to those of the two-layer case. 

%%%%%%%%%%%%%%%%%%%%%%%%%%%%%%%%%%%%%%%%%%%%%%%%%%%%%%%%%%%%%%%%%%%%%%
%%%%%%%%%%%%%%%%%%%%%%%%%%%%%%%%%%%%%%%%%%%%%%%%%%%%%%%%%%%%%%%%%%%%%%

\section{Discussion and summary}
\label{sum}

Our results might be of interest in biological applications such as
the transfer of a small molecule through a membrane from one cell to
another. Earlier theories have been developed to describe the vdW
interaction between molecules or small particles and a solvent medium
\cite{Israelachvili1974}. In particular, such particles were modeled
by a small dielectric sphere of radius $R_s$ and (macroscopic)
permittivity $\varepsilon_s$. It was found that the nonretarded
dispersion potential of such a sphere near the interface of two
dielectric media (with the sphere being situated in a medium of
permittivity $\varepsilon_2$, and $\varepsilon_1$ denoting the
permittivity of the medium on the far side of the interface) is given
by
\begin{equation}
\label{Ubio}
U(z_s)=-\frac{\hbar }
{16\pi^2 \varepsilon_0 z_s^3}\int _0^\infty \mathrm{d}\xi\; 
\frac{\alpha_s
}{\varepsilon_2}\,
\frac{\varepsilon_1-\varepsilon_2}{\varepsilon_1+\varepsilon_2}.
\end{equation}
Here, $z_s$ is the distance from the sphere to the surface and
\begin{equation}
\label{abio}
\alpha_s
(\omega)=4\pi \varepsilon_0 R_s^3\varepsilon_2(\omega)
\frac{\varepsilon_s(\omega)-\varepsilon_2(\omega)}
{\varepsilon_s(\omega)+2\varepsilon_2(\omega)}
\end{equation}
is the excess or effective polarisability \cite{McLach1965,Landau63}
of the dissolved particle in the medium. This potential already
accounts for the fact that the (macroscopic) particle has a finite
volume and can hence only move by displacing an equal volume of
solvent from its path (with associated pressure forces being present);
the excess polarisability and hence also the potential must vanish
when a dissolved particle has the same properties as the solvent.

Equation~(\ref{Ubio}) is the macroscopic counterpart of our
microscopic equations (\ref{Ubshort})--(\ref{C2}), where the
position-dependent part of our potential for purely dielectric solvent
media reads
\begin{equation}
\label{Ubshort1}
U_2(z_A)=-\frac{\hbar}{16\pi^2\varepsilon_0 z_A^3}
 \int_0^{\infty}\mathrm{d}\xi 
\frac{\alpha}{\varepsilon_2}
\left(\frac{3\varepsilon_2}{2\varepsilon_2+1}\right)^2
\frac{\varepsilon_1-\varepsilon_2}{\varepsilon_1+\varepsilon_2}.
\end{equation}
The results from a microscopic description of the particle as a
system of bound point charges, leading to our Eq.~(\ref{Ubshort1}),
are thus formally very similar to those from a macroscopic model of
the particle as a dielectric sphere, Eq.~(\ref{Ubio}). The main
difference is the fact that the microscopic
polarisability~(\ref{alpha}) together with a local-field correction
factor appears in Eq.~(\ref{Ubshort1}), while the macroscopic excess
polarisability~(\ref{abio}) which effectively accounts for pressure
forces enters Eq.~(\ref{Ubio}). Depending on the size of the immersed
particle, one of the two models may provide a more realistic
description: For very small particles like atoms or small molecules
whose size is comparable to the free interspaces between the atoms
forming the solvent medium, local-field effects are important while
(macroscopic) pressure forces can not even be defined, so that the
microscopic result~(\ref{Ubshort1}) should be used. For larger
molecules whose volume covers a region that would otherwise be
occupied by a large number of solvent atoms, local-field effects can
be neglected while pressure forces become relevant, so
Eq.~(\ref{Ubio}) should be given preference. As an additional
difference, note that our microscopic calculation has also given
layer-dependent constant contributions to the potential, which are
absent in Eq.~(\ref{Ubio}).

Similar considerations apply in the retarded limit, where the
macroscopic potential of the small sphere was found to be
\cite{Israelachvili1974}
\begin{equation}
\label{Ubio_ret}
U(z_s)=-\frac{23\hbar c}{320 \pi^2 \varepsilon_0 z_s^4}\,
\frac{\alpha_s(0)}{\varepsilon_2^{3/2}(0)}\,
\frac{\varepsilon_1(0)-\varepsilon_2(0)}
{\varepsilon_1(0)+\varepsilon_2(0)}\,;
\end{equation}
whereas our microscopic result as given by Eqs.~(\ref{Ublong}) and
(\ref{C1}) reads
\begin{multline}
\label{Ubio_ret1}
    U_2(z_A)=\frac{3\hbar c}{64\pi^2\varepsilon_0 z_A^4}\, 
    \frac{\alpha(0)}{\varepsilon_1^{3/2}(0)}
\left[\frac{3\varepsilon_2(0)}{2\varepsilon_2(0)+1}\right]^2
\\
\times\int_1^\infty\mathrm{d}y\Biggl[
\biggl(\frac{1}{y^4}-\frac{2}{y^2}\biggr)
\frac{ay-\sqrt{y^2-1+a}}{ay+\sqrt{y^2-1+a}}\\
+\frac{1}{y^4}
\frac{y-\sqrt{y^2-1+a}}{y+\sqrt{y^2-1+a}}\Biggr]
\end{multline}
[$a=\varepsilon_1(0)/\varepsilon_2(0)$]. In addition to the
observations made for the nonretarded limit, the interface-dependent
proportionality constants [i.e., the last factors in
Eqs.~(\ref{Ubio_ret}) and (\ref{Ubio_ret1})] are now also different in
general, note that they do agree in the limit of small dielectric
contrast between the media, $\chi(0)=
\varepsilon_1(0)-\varepsilon_2(0)\ll \varepsilon_2(0)$.

To conclude, we have the local-field corrected vdW potential of a
ground state atom embedded in a planar magnetodielectric multilayer
system by analytical and numerical means. The theory allows us to
extend earlier studies of the same system in two important aspects:
first, one can allow for the atom to be embedded in a medium thus
making the theory applicable to a larger range of realistic
situations; second, the effects of the local-field correction is
elucidated.

The formulas for the interaction potential have been derived for an
arbitrary number of layers where the case of two- and three-layer
systems have been studied in detail. We have shown that the potential
can be decomposed into two parts: A layer-dependent constant part
which depends on the real-cavity radius, i.e., the density of the
medium the atom is placed in; and a position-dependent part which
contains the local-field correction as a factor in the integral over
frequency. For the latter, we have presented retarded and non-retarded
limits and the considered case of neighboring media of similar
properties. Distance laws have been reestablished with effects of the
local field included where the local-field correction has been found
to be as high as 80\% in certain cases. Further, numerical
calculations show that an interplay between electric and magnetic
properties of the  neighboring media may lead to the appearance of
potential wells or walls near the surface. These structures are
potentially helpful as a trapping mechanism. Although these structures
are much less intense in magnitude than those occurring for an excited
atom \cite{Sambale2008a, Sambale2008b}, they are more permanent
because an atom in the ground state has an infinitely long lifetime.

The constant part of the potential, which originates from local-field
effects, does not contribute to the vdW forces, but it can facilitate
our understanding of the movement of an atom near an interface, and is
instrumental in our proposed estimate of the on-surface value of the
potential: After calculating the total potential in both sides of an
interface up to distances equal to the radius of the (real) cavity,
beyond which a macroscopic  model no longer applies, the average of
the two potential values taken at these distances can be regarded as
the potential at the interface. Our procedure improves previous
similar estimates by including local-field effects and consistently
removing self-interactions.

In the case of the three-layer system, emphasis was given to the
influence of the thickness of the middle layer. While new features may
arise for a very thin middle layer, the behaviour of the potential is
simply those at the interfaces combined if the middle layer is thick
enough.

Our results, which have been obtained on the basis of an exact,
microscopic model of the atom--field coupling, are complementary to
previous more macroscopic dispersion potentials of particles modelled
as small spheres. The model of preference depends on the size of the
particle in the specific situations considered. In the future, efforts
should be taken to find a dispersion potential which holds for all
possible ranges of particle sizes and includes both the microscopic
and macroscopic potentials as limiting cases. Our considerations may
easily be extended to other geometries, such as spherically or
cylindrically layered host media.

%%%%%%%%%%%%%%%%%%%%%%%%%%%%%%%%%%%%%%%%%%%%%%%%%%%%%%%%%%%%%%%%%%%%%%

\acknowledgments

The work was supported by Deutsche Forschungsgemeinschaft. S.~Y.~B.
and H.~T.~D. are grateful to the Alexander von Humboldt Stiftung for
support.

%%%%%%%%%%%%%%%%%%%%%%%%%%%%%%%%%%%%%%%%%%%%%%%%%%%%%%%%%%%%%%%%%%%%%%

\end{document}